\documentclass[conference]{IEEEtran}
\IEEEoverridecommandlockouts
\usepackage{cite}
\usepackage{amsmath,amssymb,amsfonts}
\usepackage{algorithmic}
\usepackage[pdftex]{graphicx}
\graphicspath{{../figures/}}
\usepackage{lipsum}
\ifCLASSOPTIONcompsoc
\usepackage[caption=false, font=normalsize, labelfont=sf, textfont=sf]{subfig}
\else
\usepackage[caption=false, font=footnotesize]{subfig}
\fi
\usepackage{booktabs,caption}
\usepackage[flushleft]{threeparttable}

\usepackage{textcomp}
\usepackage{xcolor}

\DeclareGraphicsExtensions{.pdf,.jpeg,.png}

\usepackage{mathabx}

\usepackage{array}
\usepackage{mdwmath}
\usepackage{mdwtab}
\usepackage{eqparbox}
\usepackage{url}
\def\BibTeX{{\rm B\kern-.05em{\sc i\kern-.025em b}\kern-.08em
    T\kern-.1667em\lower.7ex\hbox{E}\kern-.125emX}}
\begin{document}

\title{A cryogenic SRAM based arbitrary waveform generator in 14\,nm for spin qubit control\\
}

\author{Mridula Prathapan, Peter Mueller, Christian Menolfi\textsuperscript{\textsection}, Matthias Br\"{a}ndli, Marcel Kossel, Pier Andrea Francese, \\ David Heim, Maria Vittoria Oropallo, Andrea Ruffino, Cezar Zota and Thomas Morf \\

\IEEEauthorblockA{\textit{IBM Zurich Research Laboratory, R\"{u}schlikon, Switzerland.} Email: mrp@zurich.ibm.com}
}
\maketitle
\begingroup\renewcommand\thefootnote{\textsection}
\footnotetext{Presently with Cisco Systems, Inc., Thalwil, Switzerland.}
\endgroup
\begin{abstract}
Realization of qubit gate sequences require coherent microwave control pulses with programmable amplitude, duration, spacing and phase. We propose an SRAM based arbitrary waveform generator for cryogenic control of spin qubits. We demonstrate in this work, the cryogenic operation of a fully programmable radio frequency arbitrary waveform generator in 14 nm FinFET technology. The waveform sequence from a control processor can be stored in an SRAM memory array, which can be programmed in real time. The waveform pattern is converted to microwave pulses by a source-series-terminated digital to analog converter. The chip is operational at 4\,K, capable of generating an arbitrary envelope shape at the desired carrier frequency. Total power consumption of the AWG is 40-140\,mW at 4\,K, depending upon the baud rate. A wide signal band of 1-17 GHz is measured at 4\,K, while multiple qubit control can be achieved using frequency division multiplexing at an average spurious free dynamic range of 40 dB. This work paves the way to optimal qubit control and closed loop feedback control, which is necessary to achieve low latency error mitigation and correction in future quantum computing systems.
\end{abstract}


\section{Introduction}
A continuous trend in the scaling of qubit numbers in quantum systems has been established over the past few years. Realization of large scale error corrected qubit systems would demand the miniaturization and integration of different components inside the cryostat. The scaling bottleneck of quantum control systems has motivated the cryo-CMOS community to explore the prospects of cryogenic circuit design at 4\,K and below. The major challenge is to meet the reliability and performance requirements put forth by the future quantum computing systems, while consuming the least amount of power, limited by the available cooling power in the cryostat. Short and optimized control pulses are the key to realize fast qubit gates with improved gate fidelity. 
Fast gates suffer from leakage effects and additional unitary errors caused by the large bandwidth of the short control pulses \cite{b1}. Adopting techniques to increase the fidelity of short duration single qubit gates using optimized control pulses in a closed-loop fashion is crucial to build low error quantum systems \cite{b2}.

\begin{figure*}[t]
	\centering
	\subfloat[Envisioned scalable spin system\label{1a}]{%
		\includegraphics[width=0.3\linewidth]{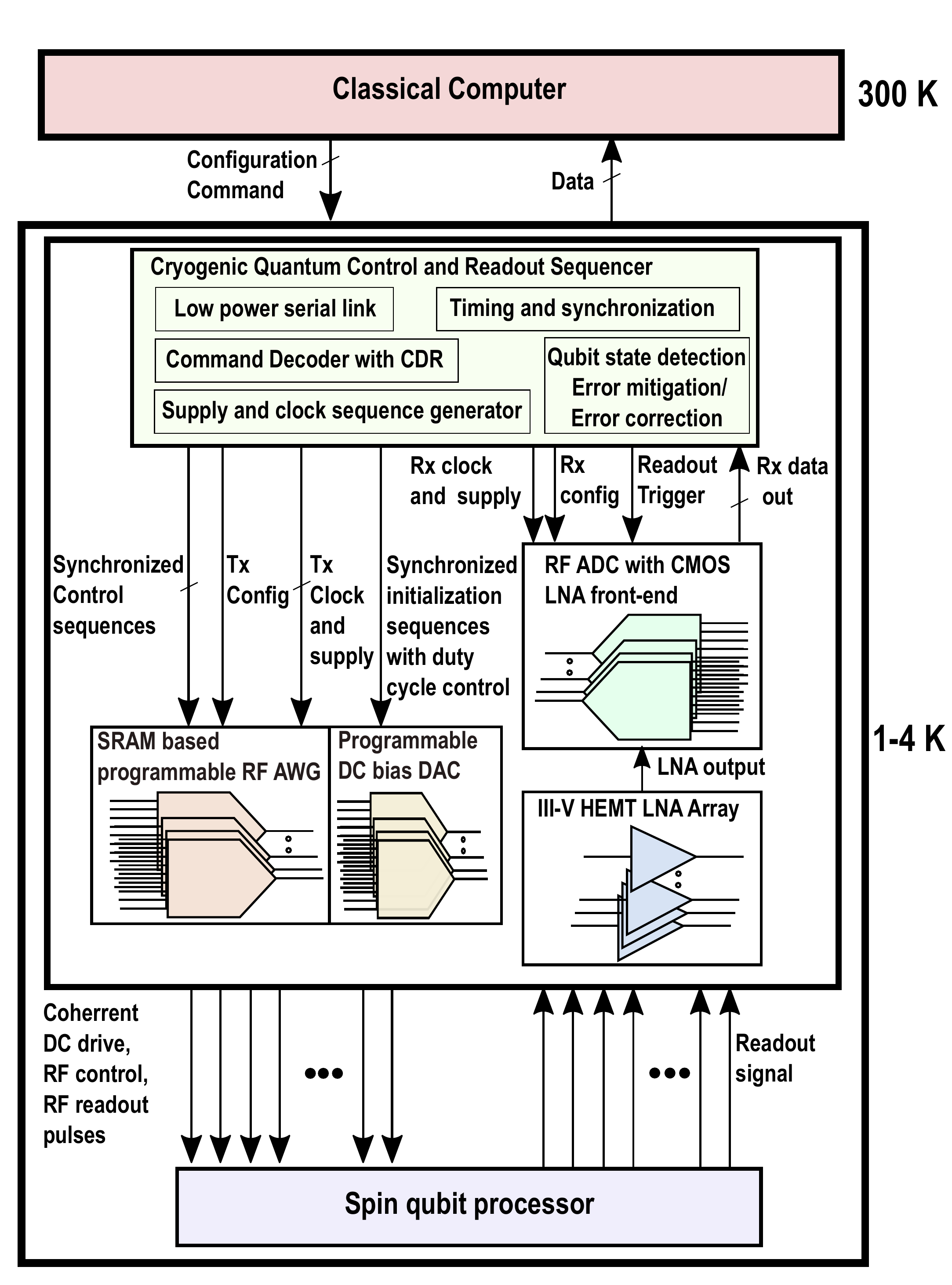}}
	\hfill
	\subfloat[RF AWG Schematic\label{1b}]{%
		\includegraphics[width=0.67\linewidth,keepaspectratio]{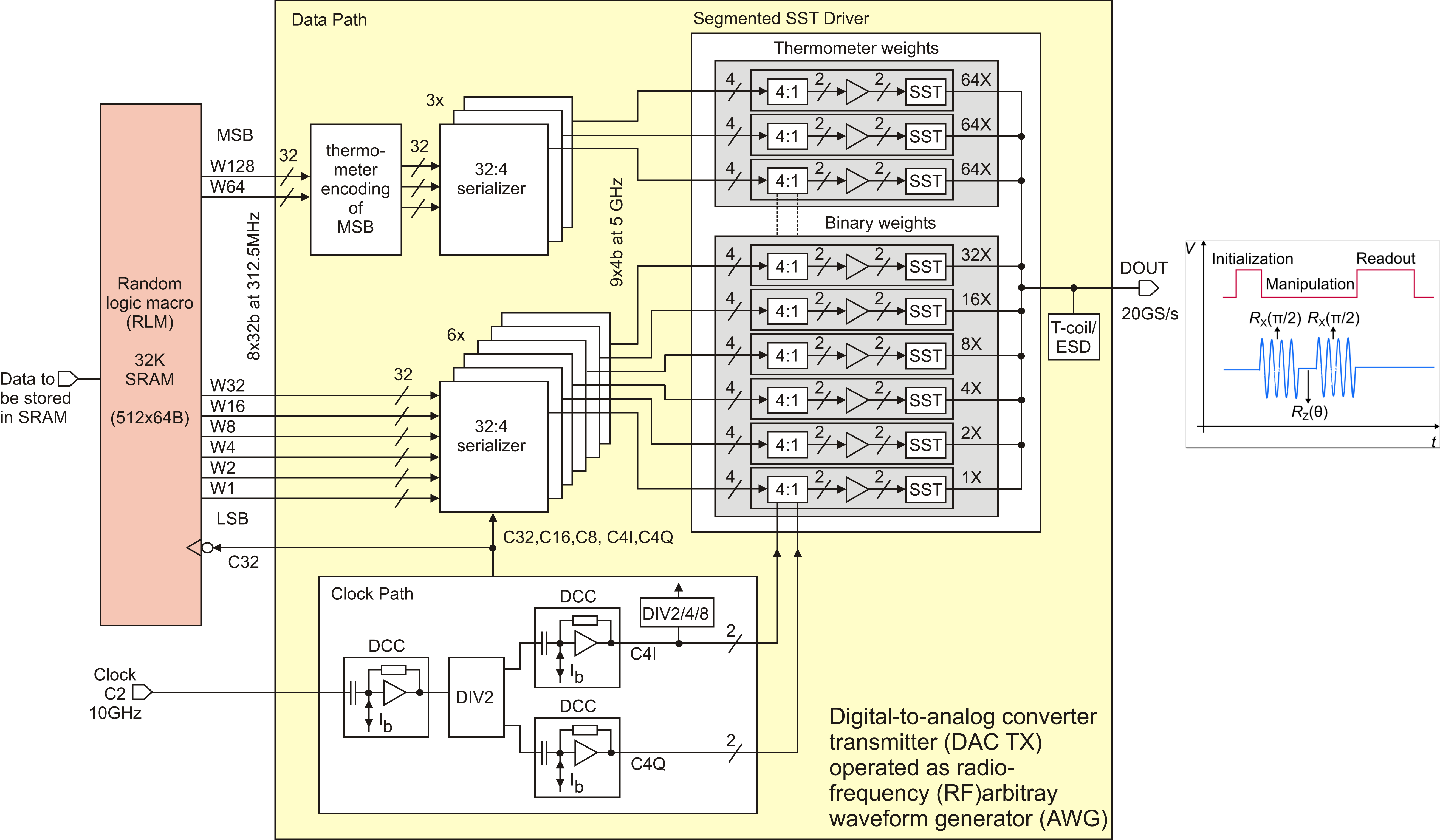}}
		\hfill
	\caption{Figure showing the SRAM based RF AWG architecture, together with the envisioned cryogenic qubit control system co-packed with hot spin qubits at 1-4\,K. The proposed AWG can be integrated into a closed-loop feedback  control system}
\end{figure*}

Spin systems have an advantage of being able to operate at higher temperatures. CMOS compatible FinFET spin qubits operating at 1-4\,K have been demonstrated by \cite{b3}. Scaling of quantum computing systems beyond 1000 qubits, irrespective of their qubit technology, puts forth an inevitable need for intelligent sub-systems operating at proximity to the qubits. These include cryogenic control and readout circuitry with maximum affordable signal processing capabilities to ensure low-latency error correction and reduced data rate for links to room temperature, as shown in Fig.~\ref{1a}. We propose an SRAM based arbitrary waveform generator, that paves the way to such systems. Advanced CMOS technology such as 14 nm bulk FinFET offers high interconnect density to support digitally assisted designs that are scalable. Control and readout circuitry implemented in advanced CMOS process nodes like 14\,nm and below could be potentially co-packaged with spin qubit arrays, a well-known advantage of spin systems over other platforms. A radio frequency (RF) digital to analog converter (DAC) with wide bandwidth and sufficient spurious free dynamic range (SFDR) could be used to control multiple qubits using frequency division multiple access (FDMA). Apart from simplifying the wiring bottleneck in the cryostat, the merits of using FDMA over 1:1 qubit to control DAC ratio for large scale error corrected systems need to be evaluated. 

The source series terminated (SST) transmitter proposed in \cite{b4} and its improved version \cite{b5} have a proven track of record in high speed I/O links applications. The advantages of SST concept are low power operation, high content of digital circuitry that supports scaling, and comparatively large signal swing. These concepts found their application in quantum control electronics owing to their scalability and overlapping specifications with high speed IO links, with design enhancements to achieve low power. The transmitter used in this work was designed by C. Menolfi et al., as a single ended adaptation of \cite{b4}. The main difference in architecture compared to the SST transmitter from \cite{b4}, is the digital part with an SRAM based pattern generator. SRAM based architectures offer high versatility with the ability to store multiple gate sequences for qubit control. Moreover, they allow the control signals to be pre-distorted, in order to compensate the non-linearities and drifts caused by cabling. It helps to mitigate the effect of intersymbol interference (ISI) caused by cabling via feed forward equalization (FFE) at no added power cost.   

\section{Chip Architecture}



The RF DAC architecture shown in Fig.~\ref{1b} is an 8-bit SRAM-based single-ended SST transmitter. Four dedicated SRAM instances, each 512$\times$16\,B provide a total 32\,KB on-chip memory. There is a 8$\times$32\,b wide interface between SRAM and DAC TX. A 1/32-rate clock, C32, is used for the data capture at the DAC TX input as well as with opposite phase for the clocking of the SRAM. Because the most significant bit (MSB) is thermometer encoded to improve the linearity and also reduce the fanout spread between MSB and least-significant bit (LSB) weights in the DAC output stages, the 8\,b resolution is implemented with nine 32:4 serializers that multiplex the input data to 9$\times$4\,b data streams at quarter-rate. Each quarter-rate data stream drives a specific DAC weight consisting of 4:1 multiplexer, pre-driver and SST output stage. The DAC weights are segmented into 6b binary weights between LSB and MSB-2 while the MSB and MSB-1 define the thermometer-encoded weights. The inclusion of the 4:1 multiplexer into each DAC segment increases the overall clock load but enables identical loading at the full data rate amongst all DAC weights, leading to superior timing jitter performance.

The clock path operates on an incoming half-rate clock, C2, that first goes into a duty-cycle correction (DCC) block followed by a frequency divider (DIV2) to produce quadrature clocks. Hence the DCC block in front of the DIV2 allows a quadrature-error correction (QEC). It is implemented as an AC-coupled trip-point-biased inverter with programmable bleeding current. Each quadrature clock is followed by a separate DCC block to adjust the duty cycle of the pertinent quadrature phase pairs. The quarter-rate quadrature signals C4I and C4Q are fed to the 4:1 multiplexers of the DAC segments as well as to the 32:4 serializers together with the sub-rate clocks C8, C16 and C32 that are derived from C4I via frequency division in the clock path. 

\begin{figure*}[t!]
	\centering
	\subfloat[Pattern generator\label{2a}]{%
	\includegraphics[width=0.4\linewidth]{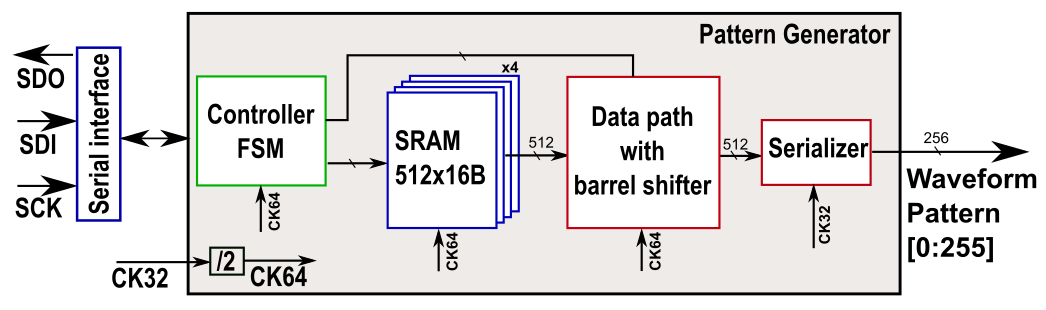}}
	\hfill
	\subfloat[Chip micrograph\label{2b}]{%
		\includegraphics[width=0.3\linewidth]{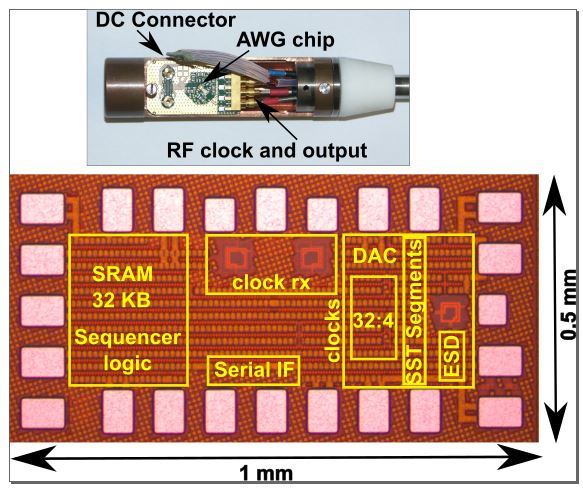}}
	\hfill
	\subfloat[Measured sequence at 4\,K\label{2c}]{%
		\includegraphics[width=0.25\linewidth]{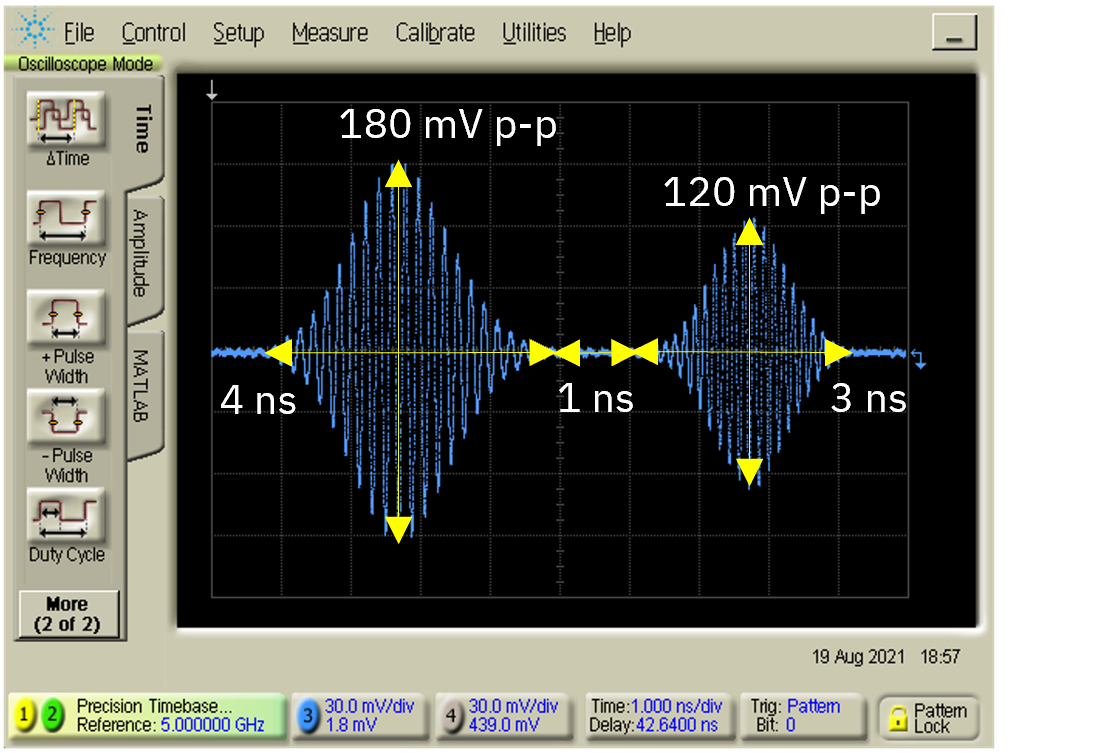}}
	\hfill
	\subfloat[Measured signal in time domain at 4\,K\label{2d}]{%
		\includegraphics[width=0.31\linewidth]{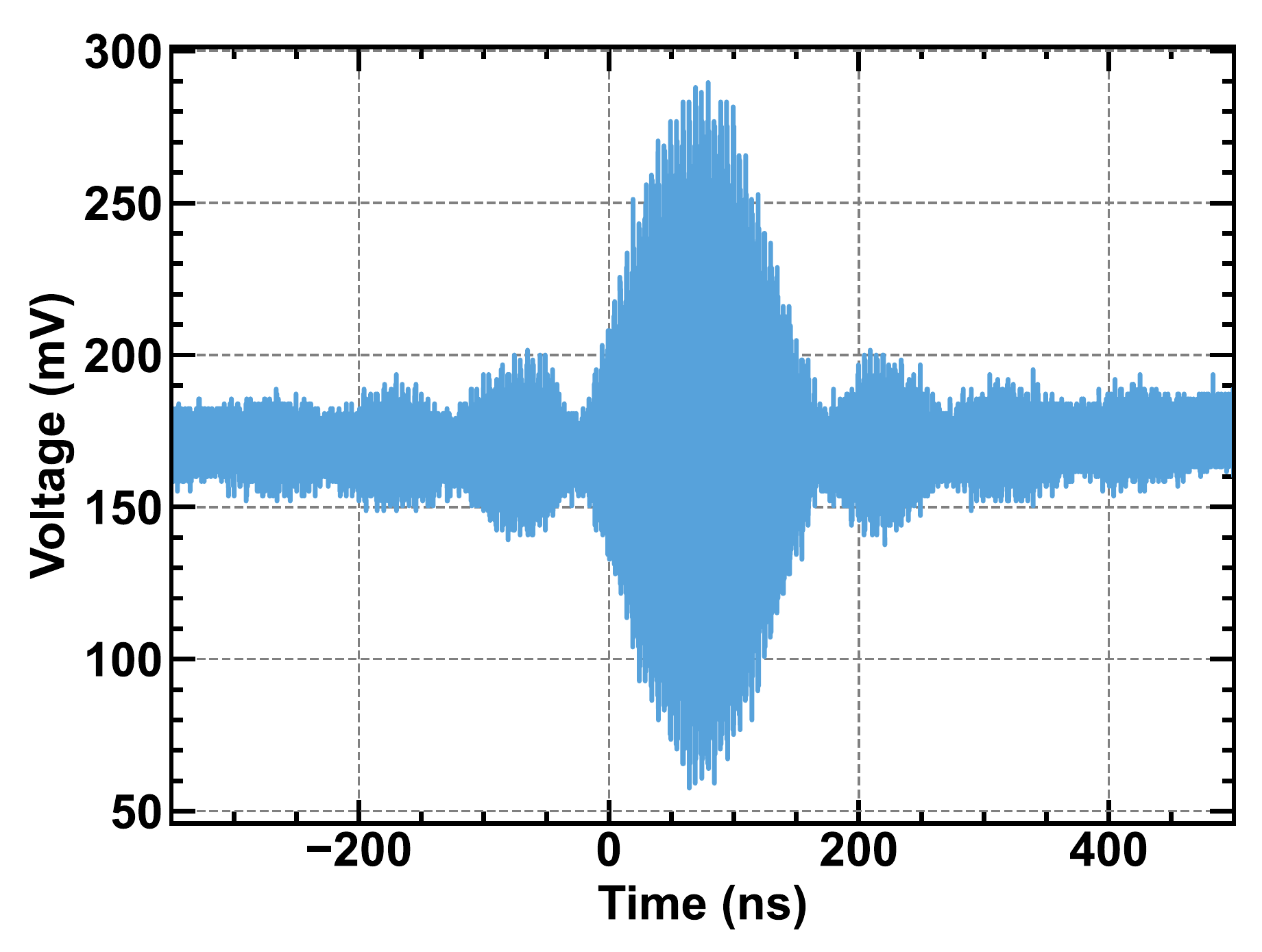}}
	\hfill
	\subfloat[Two-tone raised cosine signal measured at 4\,K \label{2e}]{%
      \includegraphics[width=0.31\linewidth]{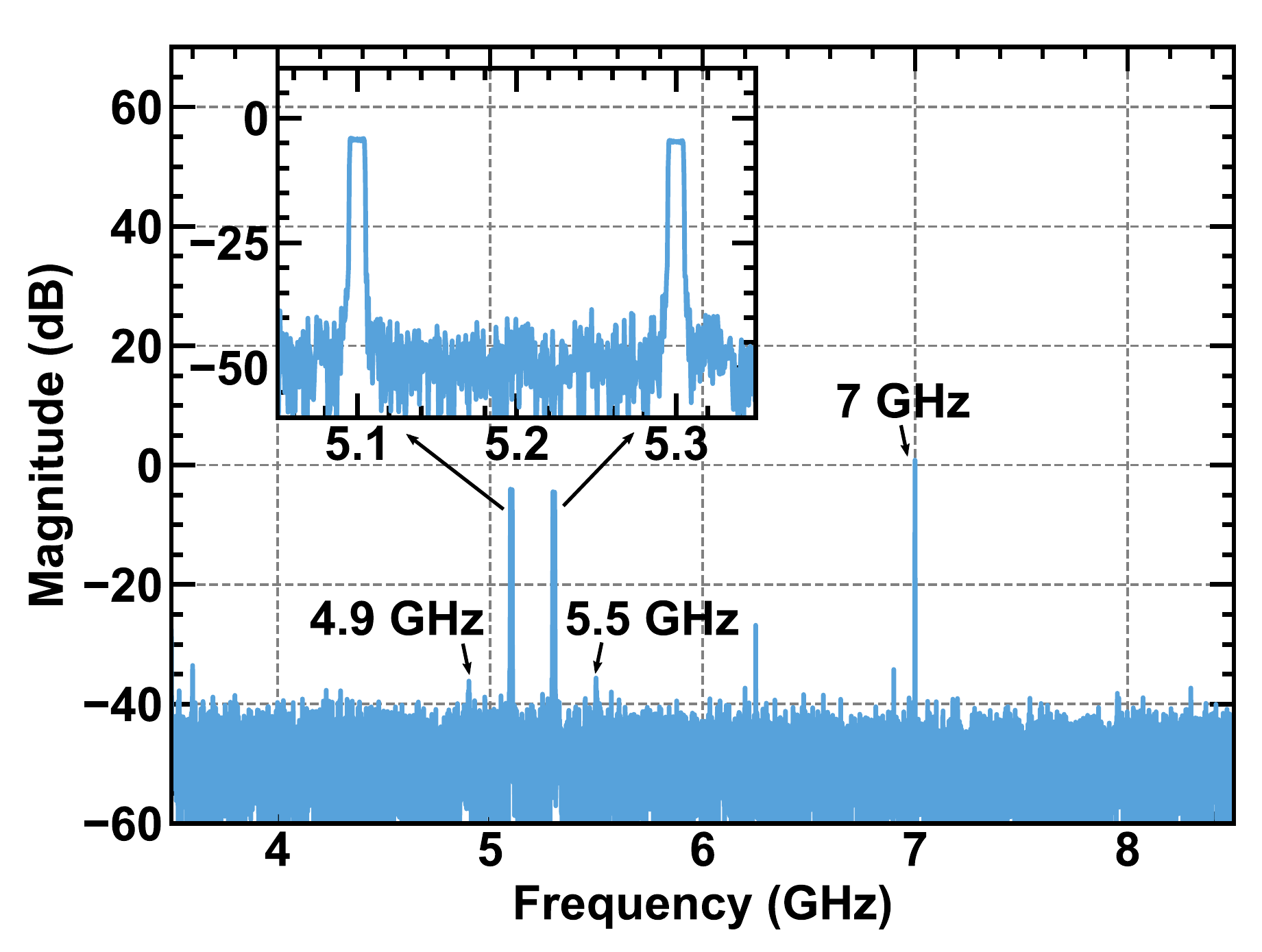}}
	\hfill
	\subfloat[Measured frequency response\label{2f}]{%
		\includegraphics[width=0.28\linewidth]{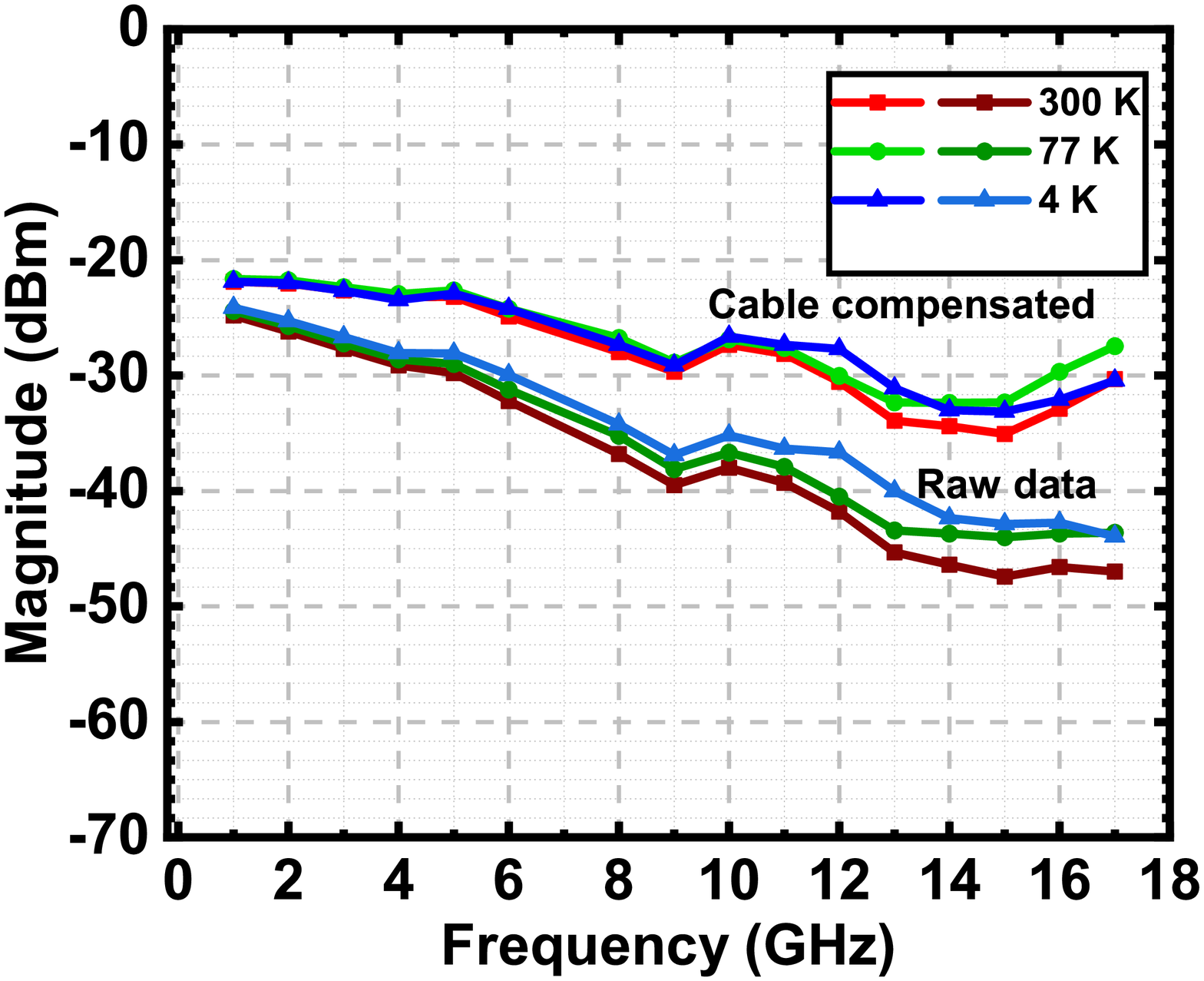}}
	\hfill
	\caption{(a) Simplified block level representation of the pattern generator (b) Chip Micrograph (bottom), dip stick setup (top) (c) Measured Gaussian sequences with programmable amplitude, pulse duration and spacing at 4\,K (d) Time domain dual-tone raise cosine signal (e) Frequency domain shows the 3\textsuperscript{rd} order intermodulation products at 4\,K for 5.1 and 5.3 GHz. The magnitude is normalized to the  largest amplitude corresponding to clock feedthrough at 7\,GHz (f) Measured frequency response}
	\label{fig2} 
\end{figure*}

The digital part consists of a pattern generator and a bidirectional serial interface as shown in Fig.~\ref{2a}. The pattern generator is responsible for generating a 256\,b input pattern to the RF DAC. It consists of a controller state machine, 32\,KB SRAM, data path logic with byte rotational capability and a digital serializer. The design is implemented in a 14\,nm bulk CMOS FinFET technology and has an active area of 0.095\,mm\textsuperscript{2}, including T-coil, ESD and SRAM.    

\section{Cryogenic Measurements}
The experiments described in this work were conducted at room temperature, liquid N\textsubscript{2} and He (300\,K, 77\,K, 4\,K) on a dip stick set up. The chip was packaged in a miniature PCB  mounted in a cryogenic dip stick shown in Fig.~\ref{2b}. The bandwidth of interest is identified as 4-8\,GHz, that matches both transmons and the spin qubits in general. The experiments target the control requirements of the spin system described in \cite{b3}. 
Fig.~\ref{2c} shows measured sub-5\,ns pulse sequences at 4\,K, showing granular control over their amplitude, pulse duration and spacing. 
At 4\,K the measured signal shows larger signal amplitude when compared to the same measurement at room temperature and 77\,K (not shown in the figure). The effect could be attributed to a combined effect of increase in the device drain saturation current for a given applied potential \cite{b6} and reduced cable losses in the dip stick. To perform linearity measurements, the DAC was programmed to output all 256 codes, creating a DC ramp in steps at the output. 1000's of samples were averaged for each DAC input code to calculate the output voltage. The measured INL at 4\,K is $< 2$\,LSB and DNL is $< 1$\,LSB. The ratio between MOS resistance and metal resistance does not appreciably vary over temperature, which makes the SST driver topology in this technology exceptionally linear over such broad temperature range, compared to the current steering DAC in \cite{b8}.

Fig.~\ref{2d} and \ref{2e} show a two tone raised cosine pulse of width 200\,ns in time domain and frequency domain respectively. A power of -33\,dBc IM3 at -43 dBm is calculated from Fig.~\ref{2e}, measured at 4\,K for a dual tone output of 5.1~\,GHz and 5.3~\,GHz. The tones are spaced 200\,MHz apart to match the qubit spacing in \cite{b3}. 
A 40 dB average spurious free dynamic range is achieved over a wide bandwidth of 1 to 17 GHz, which is in accordance with the current state of the art. The total harmonic distortion is less than 2\%. The operational frequency range allows the control of multiple spin qubits with different Larmor frequencies, based on the assumption that a large multiplexing ratio will be supported by the error corrected spin systems in future. 

\begin{figure*}[t!]
	\centering
	\subfloat[\label{3a}]{%
	\includegraphics[width=0.25\linewidth]{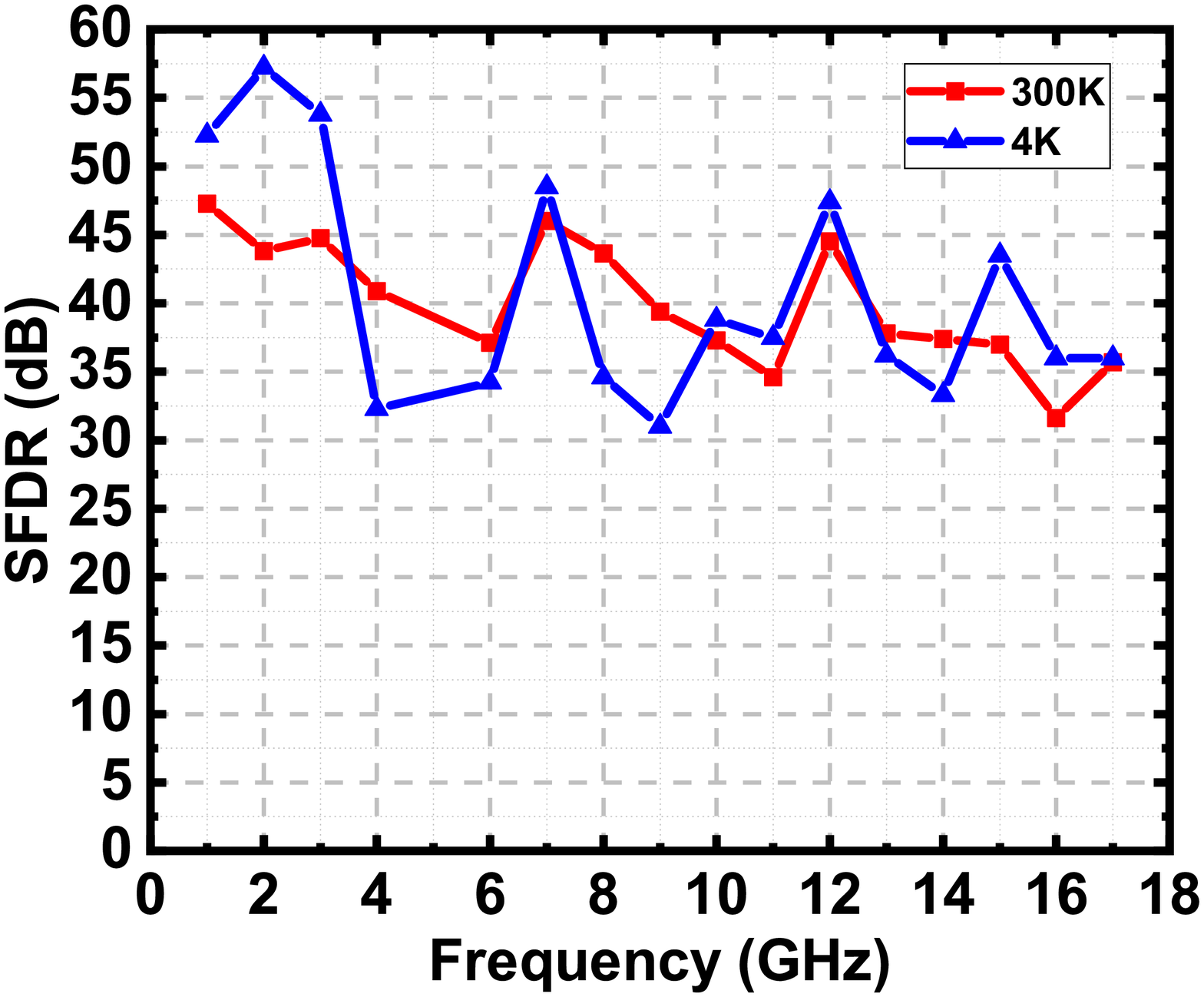}}
    \hfill
	\subfloat[\label{3b}]{%
		\includegraphics[width=0.26\linewidth]{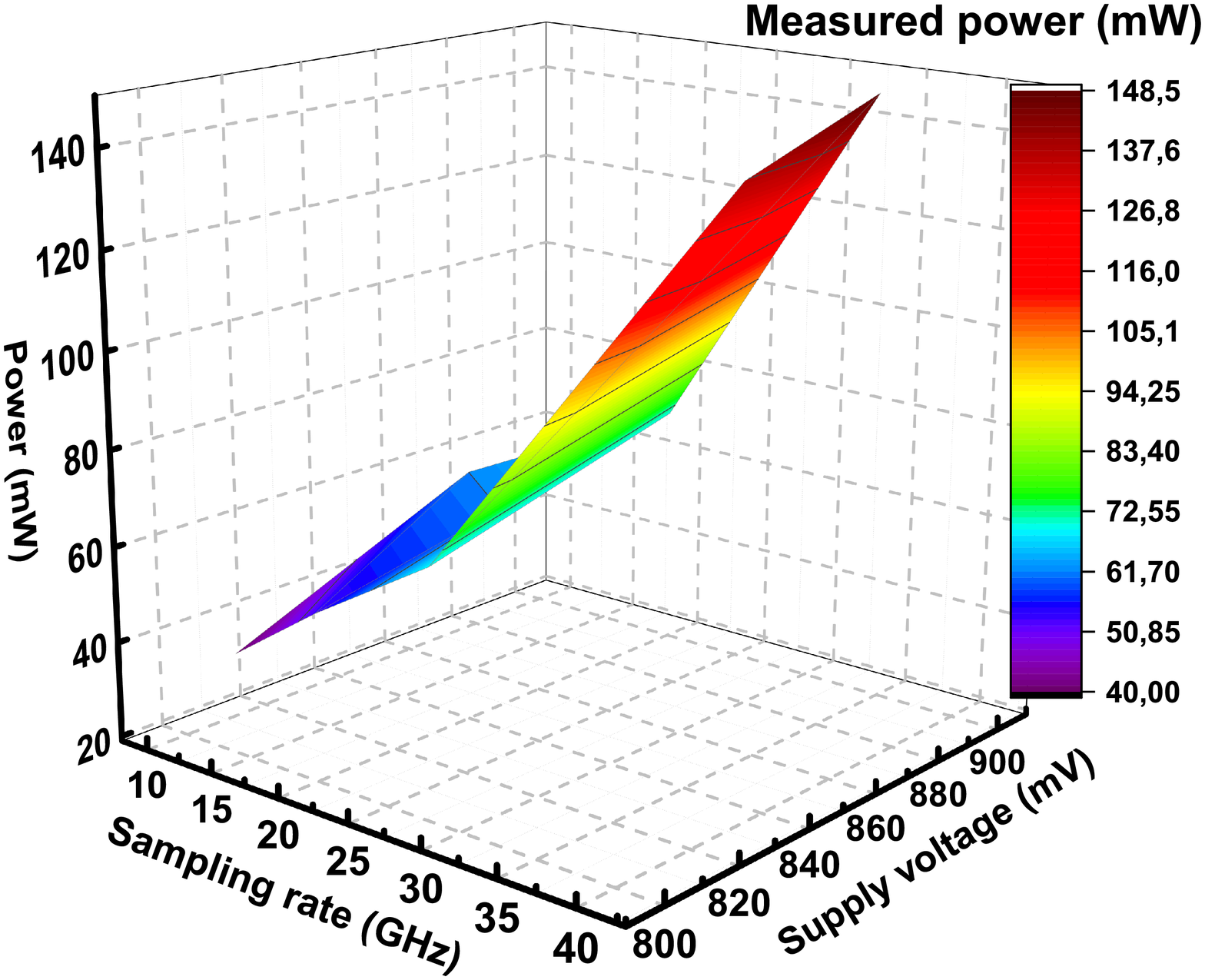}}
	\hfill
	\subfloat[\label{3c}]{%
		\includegraphics[width=0.2\linewidth]{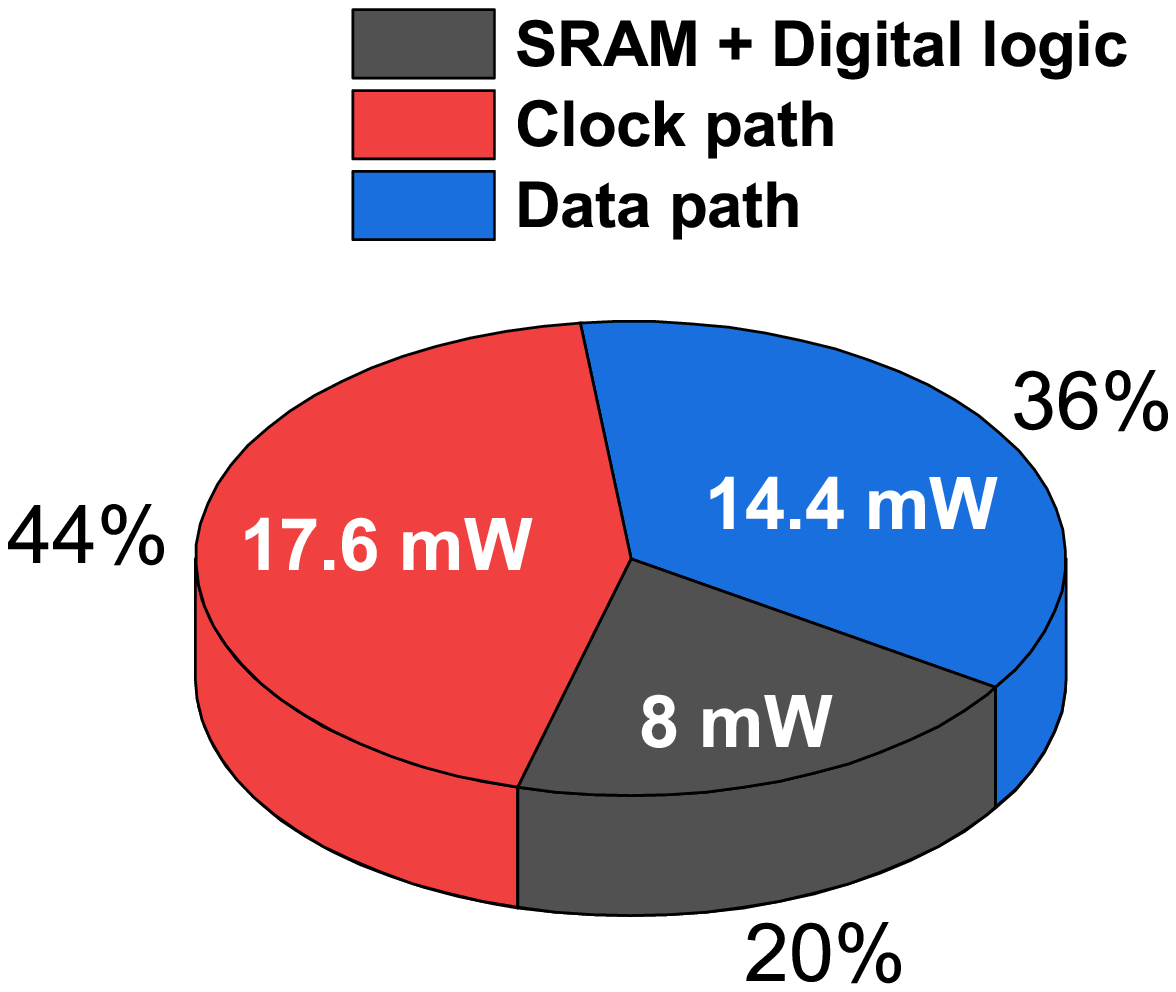}}
		\hfill
			\subfloat[\label{3d}]{%
			\includegraphics[width=0.15\linewidth]{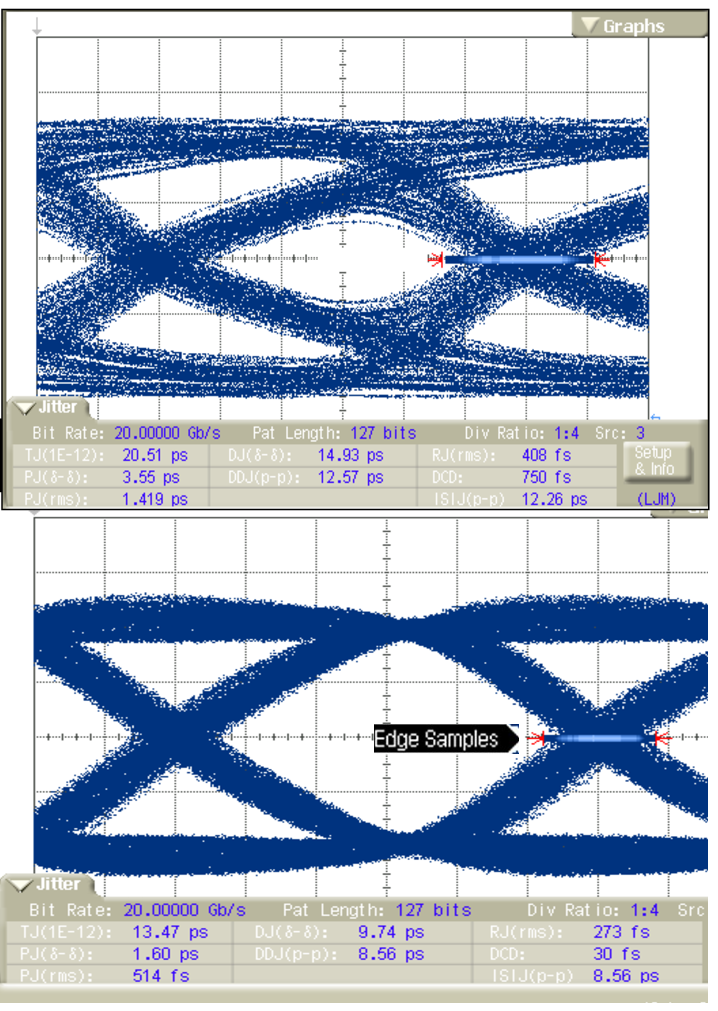}}
	\caption{(a) SFDR plots with unaccounted cable losses at 4\,K  (b) Power vs Sampling clock and Supply voltage at 4\,K. The SRAM is programmable throughout the range. (c) Power breakdown chart at 14 GHz and 0.8\,V, (d) PRBS7 eye diagram at 20 GS/s at 4\,K shows that the deterministic jitter due to cabling improves by 35\% by applying feed forward equalization}
	\label{fig3} 
\end{figure*}

Jitter measurements were performed to analyze the drift on the control signal. Optimal linear equalizer coefficients for the cryogenic set up were calculated for a known pattern such as PRBS7. A 35\% reduction in deterministic jitter is observed after feed forward equivalization (FFE) Fig.~\ref{3d}. A square wave output at 5\,GHz, with FFE, shows $<$ 2\,ps deterministic jitter at 4\,K as shown in Table~\ref{tab1}. One of the advantages of SRAM based AWG architecture lies in its ability to adapt the FFE coefficients to equalize the changes in cryostat cabling over time. Since it is done through input data pattern, no extra power is required. Moreover, it facilitates real time calibration.  

Fig.~\ref{3b} shows the power measurements at 4\,K. The SRAM is programmable throughout supply and frequency range shown in Fig.~{3b}. The power consumption could be further minimized by dynamically lowering the power supply of the SRAM, while the data is held safe. Although the digital part of the design operates at a fraction of the reference clock, the total power consumption increases with the sampling frequency (2$\times$reference clock), as shown in Fig.~\ref{3b}.  
On the other hand, high frequency operation of RF up conversion architectures \cite{b7}-\cite{b9} would demand high frequency LO and mixer, adding up to their total power consumption. The digital blocks consume only 20\% of the total power, where as the power consumption of \cite{b8}-\cite{b9} is largely dominated by the digital part as shown in Table~\ref{tab2}. The total power consumption is often reported as power per qubit for a given number of frequency multiplexed qubits per controller. This ratio may vary depending upon the DAC bits, linearity, target SFDR, qubit frequency spacing, etc. A total power of 2-4\,mW/qubit can be estimated for the proposed controller, with a multiplexing ratio of 1:20 (drive:qubit) in our target qubit platform \cite{b3}, with the qubit frequency spacing of 200\,MHz within 4-8\,GHz band.      
\begin{table}[tbh]
	\caption{Jitter measurements for square wave at 5 GHz}
	\begin{center}
		\centering
		\begin{tabular}{|c|c|c|c|}
			\hline
			\textbf{Temperature} & \textbf{\textit{Total jitter$^{\mathrm{a}}$}}& \textbf{\textit{Random$^{\mathrm{a}}$}}& \textbf{\textit{Deterministic$^{\mathrm{a}}$}} \\
			\hline
			300\,K & 4.80 ps & 0.237 ps& 1.56 ps \\
			\hline
			77\,K& 5.91 ps & 0.239 ps & 2.64 ps \\
			\hline
			4\,K& 4.98 ps & 0.221 ps & 1.95 ps \\
			\hline
			\multicolumn{4}{l}{$^{\mathrm{a}}$Measurements were done using precision timebase scope.}
		\end{tabular}
		\label{tab1}
	\end{center}
\vspace{-7mm}
\end{table}
%


\begin{table}[b!]
	\caption{Comparison with current state of the art}
	\begin{center}
		\centering
		\begin{tabular}{|p{0.54in}|p{0.54in}|p{0.54in}|p{0.54in}|p{0.45in}|}
			\hline
			\textbf{Features} & \textbf{\textit{This work}}& \textbf{\textit{\cite{b9}}}& \textbf{\textit{\cite{b8}}} & \textbf{\textit{\cite{b7}}} \\
			\hline
			Qubit \mbox{Platforms} & spin & spin & transmons, spin& transmons \\
			\hline
			Functionality & RF control & Gate pulsing, RF control, Readout & RF control & RF control\\
			\hline
			Chip area & 0.5\,mm\textsuperscript{2} & 16\,mm\textsuperscript{2} & 4\,mm\textsuperscript{2}& 1.6\,mm\textsuperscript{2} \\
			\hline
			Technology& 14\,nm\,FinFET & 22\,nm\,FinFET & 22\,nm\,FinFET & 28\,nm\,bulk\\
			\hline
			Power \mbox{consumption} & Total\,power: 40-140\,mW Analog: 32-54\,mW Digital: 8-16\,mW & Total drive\,power$^{\mathrm{b}}$: 93-223\,mW  Analog: 83\,mW or 5.2\,mW/qubit Digital: 10-140\,mW & Total\,power$^{\mathrm{b}}$: 384.4\,mW Analog: 54.4\,mW or 1.7\,mW/qubit Digital: 330\,mW & 2\,mW/qubit \\ 
			\hline
			Frequency multiplexing & yes & yes & yes & no\\
			\hline
			\mbox{Drive to} \mbox{qubit ratio} & N/A & 1:16 & 1:32 & N/A\\
			\hline
			Frequency range & 1-17 GHz & 11-17 GHz & 2-20 GHz & 4-8 GHz \\
			\hline
			IM3 & $-33$\,dBc at 
			$-43$\,dBm & $-50$\,dBc at $-17$\,dBm & $-50$\,dBc at $-18$\,dBm & N/A \\
			\hline
			On-chip storage & 32\,KB SRAM & RAM, \mbox{digital} memory & SRAM & Digital memory \\
			\hline
			Waveform storage & 32\,K\,points AWG & 16\,K\,points AWG & 40\,K points AWG & Fixed 22\,points \\
			\hline
			DAC bits & 8 & 10 & 10 & 11 \\
			\hline
			\end{tabular}
		\begin{tablenotes}
			\small
			\item {$^{\mathrm{b}}$ Calculated by multplying the reported power/qubit by the reported number of qubits per controller for the given system}
			\end{tablenotes}
		\label{tab2}
	\end{center}
\end{table}
\section*{Conclusions}
A study on the cryogenic operation of an SRAM based AWG with an RF DAC in 14 nm FinFET technology is reported. The proposed AWG can generate RF control pulses of any envelope shape with programmable amplitude, carrier frequency, pulse spacing and duration. Furthermore, it enables real time feed forward equalization to compensate drift due to changes in the cryostat cabling, at no added power cost. The SRAM is fully functional and programmable at 4\,K, within the given voltage and frequency range. The proposed digitally dominated architecture eliminates the need of local oscillators and mixers, reaping the benefits of scaling in advanced CMOS nodes, such as 14\,nm or below. The measured wide signal band of 1\,GHz to 17\,GHz at 4\,K is in accordance with the specifications of transmons and spin qubits developed thus far.
Mutiple qubit control can be acheived via frequency multplexing with a target average spurious free dynamic range of 40\,dB. 
The proposed AWG architecture supports system level integration of a cryogenic signal processor, leading to fully integrated, feedback based qubit control systems at 4\,K.  


 
\section*{Acknowledgment}
This work was supported as a part of NCCR SPIN, a National Centre of Competence in Research, funded by the Swiss National Science Foundation (grant number 51NF40-180604). The authors thank Ralph Heller, Daniele Caimi, and BRNC for the technical support.


\begin{thebibliography}{00}
\bibitem{b1} Motzoi, F., Gambetta, J. M., Rebentrost, P., Wilhelm, F. K. Simple pulses for elimination of leakage in weakly nonlinear qubits. Phys. Rev. Lett. 103, 110501 (2009).
\bibitem{b2} Werninghaus, M., Egger, D.J., Roy, F. et al. Leakage reduction in fast superconducting qubit gates via optimal control. npj Quantum Inf 2021. 
\bibitem{b3} Camenzind, L. C., Geyer, S., Fuhrer, A. et al. A hole spin qubit in a fin field effect transistor above 4 kelvin. Nat Electron (2022)
\bibitem{b4} C. Menolfi et al., "A 112Gb/S 2.6pJ/b 8-Tap FFE PAM-4 SST TX in 14nm CMOS," 2018 ISSCC, 2018,doi: 10.1109/ISSCC.2018.8310205.
\bibitem{b5} M. A. Kossel et al., "8.3 An 8b DAC-Based SST TX Using Metal Gate Resistors with 1.4pJ/b Efficiency at 112Gb/s PAM-4 and 8-Tap FFE in 7nm CMOS," ISSCC, 2021, pp. 130-132
\bibitem{b6} A. Chabane et al., "Cryogenic Characterization and Modeling of 14 nm Bulk FinFET Technology," ESSDERC-ESSCIRC, 2021, pp. 67-70
\bibitem{b7} J. C. Bardin et al., “Design and Characterization of a 28-nm Bulk-CMOS Cryogenic Quantum Controller Dissipating Less Than 2 mW at 3 K,” IEEE Journal of Solid-State Circuits, vol. 54, no. 11, Nov. 2019.
\bibitem{b8} J. P. G. Van Dijk et al., "A Scalable Cryo-CMOS Controller for the Wideband Frequency-Multiplexed Control of Spin Qubits and Transmons," in IEEE Journal of Solid-State Circuits, vol. 55, no. 11, pp. 2930-2946, Nov. 2020, doi: 10.1109/JSSC.2020.3024678.
\bibitem{b9} Park et al., "A Fully Integrated Cryo-CMOS SoC for State Manipulation, Readout, and High-Speed Gate Pulsing of Spin Qubits," in IEEE Journal of Solid-State Circuits, Nov. 2021, doi: 10.1109/JSSC.2021.3115988.
\end{thebibliography}
\end{document}